\documentclass{article}

\usepackage{amsmath}  
\usepackage{amsfonts} 
\usepackage{graphicx} 
\usepackage{textcomp}
\usepackage{indentfirst}
\usepackage[authoryear]{natbib}
\bibpunct{(}{)}{;}{a}{}{,}
\begin{document}


\title{Cosmological redshift, recession velocities and acceleration measures in FRW
cosmologies}

\author{Alexey V. Toporensky, Sergei B. Popov}





\maketitle 

\noindent
{\bf Abstract}\\
In this methodological note we discuss several topics related to interpretation of some
basic cosmological principles. We demonstrate that one of the key points is
the usage of synchronous reference frames. The Friedmann-Robertson-Walker
one
is the most well known example of them. We describe how different
quantities behave in this frame. Special attention is paid to potentially
observable parameters.  
We discuss different variants for choosing measures of velocity and acceleration representing the Hubble
flow, and present illustrative calculations of apparent acceleration in flat
$\Lambda CDM$ model for various epochs.  
We generalize description of the ``tethered'' galaxies problem
for different velocity measures and equations of state, and illustrate time
behavior of velocities and redshifts  in the $\Lambda CDM$
model.  

\section{Introduction}
 There are many controversial issues related to interpretations of
basic principles in cosmology. In addition, there are several widespread 
misconceptions (see discussions about many of them in 
\citealt{davis2005phd}, and references therein) and some
examples of inappropriate choice of parameters to illustrate  different
aspects of cosmological phenomenae. Most of these complications are 
due to the fact that cosmological observables are frame-dependent, which
often is not fully acknowledged. The choice of a frame dictates which
distances, velocities etc. fit better and do not mislead the
discussion. Even disputes about the interpretation of the cosmological
redshift (if it should be treated as some ``third type'' of redshift
related to the expansion of space, or it can be reduced to the
well-known types of redshift) can be, at least partly, reduced to the
discussion about frames.

 Many papers have been dedicated to discussions of these
misconceptions and difficulties. Without pretending to give a complete list
we want to note several of them.
 The problems of superluminal velocities and many related issues have been
discussed by \cite{davis2004} (see also references therein for earlier
results, among which we especially recommend  \citealt{murdoch1977}).
 Interpretation of the cosmological redshift was the main topic of
numerous studies. In particular, we want to mention \cite{kaiser2014}
(references to earlier papers can be found there).

 This paper is a continuation of the discussion started in
\cite{tp2014}. In the following sections we duscuss the role of
the synchronous nature of the Friedmann-Robertson-Walker (FRW) metric,
discuss cosmological redshifts, describe four different measures for the Hubble
flow, and
discuss their advantages and disadvantages,
then we discuss how time derivatives of the redshift and of the velocity at the moment
of emission evolve in the $\Lambda CDM$ cosmology,
and finally, we 
illustrate the behavior of so-called ``tethered'' (see,
\citealt{davis2003}) galaxies in different cosmological models. 
All the main parameters used
in our discussion are summarized in the Table at the end of the paper.

\section{Light propagation and time intervals in the FRW metric}

The metric in the FRW frame has the well-known form:
$$
ds^2=dt^2-a^2(t)d\chi^2-\chi^2d\Omega^2,
$$
where $t$ is the cosmic time, $a$ is the scale factor, and $\chi$ is the radial comoving coordinate.
$\Omega$ includes all angular dependences which are not discussed below as
we focus on radial motions.

As the coordinate system is synchronous, all points forming it share the same proper time.
There is also an agreement about values of spatial proper distances which are equal to $a(t)
(\chi_1-\chi_2)$ for all observers despite their relative motion.

Note that points are moving with respect to each other --- 
distances are changing with the rate $v=\dot a (\chi_1-\chi_2)$. This represents
the existence of the Hubble flow: points recede from each other due an
increase in the scale factor $a$, while the comoving coordinate remains unchanged.
As we will see soon, these velocities of the Hubble flow (recession velocities) have particular
properties in the FRW frame, and should not be confused with velocities arising from
changes of the comoving coordinate (peculiar velocities). 

Due to synchronous nature of the FRW frame, motion in the Hubble flow does not lead to 
any change in the proper time and distance intervals. This situation 
is possible only in the presence of a gravitational field.\footnote{Without
gravity in a  synchronous metric --- i.e. in the Minkowski space --- distances are not
changing. Conversely, in the presense of a gravitational field a synchronous
metric cannot be stationary. The FRW frame is the best example of such a situation.} 
This is already enough to understand why this velocity, $v$, can exceed
the speed of light --- there is no Lorentz time dilation between points in the Hubble flow,
so we can not expect any limiting velocity. This is not in any contradiction
with Special Relativity (SR)
because $v$ is not a directly observable variable. 

The existence of superluminal motion in
the Hubble
flow is, as it is now well understood, a frame dependent phenomenon. We can refer to
\cite{chodor2007}
where a detailed investigation of this problem is given. 
As the author says, ``superluminarity
of distant galaxies vanishes in suitable coordinate system''. However, several paragraphs above
this phrase Chodorowski recognizes that 
``the RW coordinate is more convenient for calculation'' because it keeps
the homogeneity of the Universe while the former coordinate (in which the
superluminarity is absent)
does not. So, the practical question is as follows: 
is this superluminarity so unpleasant that it is
reasonable to avoid it at the price of more complicated calculations and losing spatial
homogeneity? The answer from a practical cosmology point of view is definitely ``no'' --
the FRW coordinates are used
everywhere they can be used. The goal of this paper is to describe what we
face in the FRW coordinates, and how to work with it correctly.

The fact that the time $t$ is the proper time for each particle in the Hubble
flow, leads 
to even worse ``blasphemy'' than the superluminarity. Another strong
deviation from a ``normal'' behavior of a system in SR appears. 
Suppose that the comoving coordinate of  a galaxy
is changing, i.e. the galaxy has a non-zero peculiar velocity $v_{p}$. 
This velocity should be added to
the Hubble flow velocity $v$ using the Galilean rule: $v_{tot}=v+v_{p}$, independently
of how large these velocities are. Indeed, we have: 
$$
d(a\chi)/dt=\chi da/dt+a d\chi/dt,
$$ 
where the first term
in the right hand side (r.h.s. hereafter) 
is the velocity of the Hubble flow, and the second term is the peculiar velocity.
Obviously,
expressions like $c/2 + c/2 =c$ may appear shocking for those who are familiar with
SR (only!) from a kindergarten. However, this is a correct way of velocity
summation
{\it if one 
of velocities is caused by the Hubble flow} (two peculiar velocities evidently obey
the relativistic law). This is true also for the speed of light which is equal to $c$ only locally.
Indeed, the equation of motion for light rays $ds^2 = 0$ leads to $d\chi/dt=\pm c/a$. Going from
comoving to physical distance we get $d(a\chi)/dt=\chi da/dt+ a d\chi/dt= v \pm c$ where two signs
indicate two possible directions of light: towards an observer, and away from it.  Again, this
is not 
a danger for SR because the speed of light is constant only in inertial frame,
which is not the case of FRW frame, being ``a hybrid of distances measured in different inertial 
frames'' (\citealt{Chodorowski:2006rb}).
 
Therefore,  a reader can treat the Hubble flow as a ``gravitational wind'' which drags objects with
locally measured peculiar velocities. It is necessary to keep in mind, however, that this 
``dragging'' is only kinematical -- there are no additional forces acting as a result of
the Hubble flow
(see \cite{davis2003, davis2004} for a detailed description). 
This picture, being rather weird from the viewpoint
of SR, may be rather comprehensive and helps to understand some unusual phenomena
specific to cosmology. 
In the next sections 
we show how some known classical cosmological results can be incorporated in this picture.

\section{Redshifts and distances}

In this section we discuss cosmological redshifts, distances, and the necessity
of the ``expansion of space'' as an additional concept to interpret 
cosmological data.

\subsection{Interpretation of the cosmological redshift}

Before considering redshifts in cosmology we briefly remind the reader
what happens in SR. A Doppler shift appears due to, in fact, two reasons. 
The first reason is purely geometrical and exists in classical physics as well.
When an emitter moves radially towards or from us, two light signals separated by
a time interval $\Delta t_{em}$ are also separated by the distance $v \, \Delta
t_{em}$. This  means
that an observer at rest would see them separated by 
a different time interval $\Delta t_{obs}=\Delta t_{em}(1+v/c)$.
This results in a classical (or kinematical) redshift $(1+z_{cl})=1+v/c$.

The second reason is related to the Lorentz transformation linking time intervals in
observer's
and emitter's frames: $\Delta t_{obs}=\Delta t_{em}/(\sqrt{1-v^2/c^2})$, 
resulting in the Lorentz part of redshift
$(1+z_L)=1/\sqrt{1-v^2/c^2}$. For a tangent motion when the first effect is absent, the Lorentz shift
is the only reason for a redshift: $z_t=z_L$. While for a radial motion the resulting 
redshift is a combination of these two effects: $(1+z_r)=(1+z_L)(1+z_{cl})$. The latter
formula usually is written in the form $z_r=\sqrt{(1+v/c)/(1-v/c)}$ where classical and Lorentz parts are not
separated. 
 In observational astrophysics Lorentzian effects are important,
for example, in GRB physics: the time of variability in the observer's
frame is much longer than in the frame of rapidly moving shells in the jet
of the burst.

The classical redshift $z_{cl}$ is responsible for another effect related to
a non-zero emitter's velocity.
Namely, it causes a difference between true and apparent velocities of the emitter seen by
an observer. Usually this effect is illustrated in a science fiction style.
Suppose, a spaceship is
traveling with a subluminal velocity $v=(4/5)c$ from $\alpha$~Cen,
located at  $\sim$4 light years from
the Earth. It starts at the moment $t_{em}$, and reaches the Earth 5 years later
according to the clocks at the terrestrial frame. 
However, a terrestrial observer
would actually 
see the departure at $t=t_{em}+4$~yrs and arrival at $t=t_{em}+5$~yrs. So, the travel apparently 
has a duration of only one year, and the apparent velocity appears to be $4c$.

This situation is well-known in astrophysics in the case of so-called superluminal
jets. The effect was predicted by \cite{rees1966}, and few years later discovered by
\cite{gubbay1969} for a near-by quasar. A detailed explanation of jet properties 
(including superluminal motions in Galactic and extragalactic sources) can be found in
the volume edited by \cite{jets2010}. The apparent velocity in this case can
be calculated as:

\begin{equation}
v_{app}=v \sin \theta/(1+\beta \cos \theta),
\end{equation}
where $\theta$ is the angle between the jet and the line of sight, and
$\beta=v/c$, $v$ is the real velocity in the jet.

As the true and apparent velocities are studied in the common frame --- the frame of the observer,
---
the effect is caused only by the difference in the duration of the time intervals at
the moments of emission and
observation, so that only classical part of the Doppler shift appears in the resulting formula. For a pure radial motion
it  takes very simple form 
 $v_{app}=v/(1+z_{cl})$. So, the apparent velocity of
an approaching emitter is larger
than the true velocity, and actually can be arbitrarily large. 
While receding objects have a smaller apparent
velocity which cannot exceed $c/2$ in special relativity.

Now having all this in mind, let us try to understand the nature of cosmological redshifts in
the FRW frame.
Suppose, an observer detects light from an object which recedes in the
Hubble flow with the velocity $v_{em}$ at the
time of emission.
As the Hubble flow ``drags'' the emitted light, so that
its velocity with respect to an observer is $c-v$, the kinematical part  of
the Doppler effect is absent at $t_{em}$:
the distance between two light pulses is equal to $c \Delta t_{em}$ instead of
$(c+v_{em}) \Delta t_{em}$ in
SR. Also, as the time coordinate is the same for any object in the Hubble flow,
no Lorentz shift is present as well. 

A difference in frequency appears due to the presence of a gravitational
field, because metric is not stationary due to gravity. In the
FRW frame two light pulses located at points with different coordinates $\chi$
would feel different
Hubble flow ``drag''. Initially, they have radial coordinate difference, $\Delta r$,
 equal to $c \Delta t_{em}$.
This difference results in the velocity difference $\Delta v = H \Delta r$ which allows us to
construct a differential equation ${d \Delta r}/{dt}=H \Delta r$. After obvious calculations
using the definition of the Hubble parameter --- $H=\dot a/ a$, --- 
we obtain that the ratio of time intervals
at $t_{em}$  and the moment of observation, $t_{obs}$, is equal to the
inverse ratio of scale factors
at these moments: 
$\Delta t_{em}/ \Delta t_{obs} = a_{obs}/a_{em}$. 
This is the well-known result for a cosmological redshift. 
A common
feature with the classical redshift $z_{cl}$ is that both have their origin in
the difference
between $\Delta t_{em}$ and $\Delta t_{obs}$ which 
have a pure geometrical nature and is calculated in the same frame.
An important difference is that a cosmological redshift is ``formed''  on
the way of the light from
an emitter to an observer, and so is not directly related to the velocity of
the emitter (we can
even construct a situation in an oscillating Universe where an object can be cosmologically
blueshifted despite being receding at the time of emission, or vice versa).
As an additional result, we obtain that the apparent velocity of an emitter in
the Hubble flow
is related to the velocity at the moment of emission as:
 
$$
v_{app}=v_{em}(\Delta t_{em}/\Delta t_{obs})=v_{em}/(1+z).
$$    

Now the question arises: how do we interpret the resulting formula for a cosmological redshift?
A widespread opinion exists, that redshifts of objects in the Hubble flow represent
some third type 
of redshift (i.e., it is different from kinematical and gravitational
redshifts) appearing due to the stretching of space.
The fact that this redshift can be defined as a ratio of scale factors 
at the moments of emission and observation seems to
support this view. Sometimes, it is claimed that the  derivation presented
above (if re-interpreted accordingly)
reduces a cosmological redshift to the Doppler effect integrated along the trajectory of the light
(see, for example, \citealt{zn1967}).
This might be rather puzzling because other types of redshift do not require any integration
along the light traveling path. And again, a reasonable explanation invokes
the concept of expanding
space with a specified rate of expansion which is integrated over all points
that the light
passes through. 

From our point of view, 
the concept of space expansion, formally, is not necessary to explain cosmological
redshifts. First of all, it is worth noting that in an inhomogeneous
Universe, in general, 
the value $1+z$ does not coincide with the ratio of scale factors (see, for
example, \citealt{Mustapha:1997xb, Basett:2013qqa}). 
Second, in our considerations
we never used the concept of expanding space. Moreover, we did not use any
specific cosmological nature of elements of
the picture under consideration. All we use is the fact that we work in a synchronous reference frame.
Such frame can be used in absolutely different physical situations. For example, in describing 
a free fall into a black hole (in this situation this role is played by
Lemaitre coordinates).
Sometimes such a redshift is refered as a tidal one. It is caused by a non-zero gravitational 
field, but in another way than the usual gravitational redshift caused by time
dilation. 

 We mentioned the Lemaitre coordinates and the tidal redshift.  It is well-known that
by using another set of coordinates (an obvious example is the Schwarzschild stationary
system of coordinates) it is possible to express the tidal redshift with the standard kinematic and
gravitational redshifts. 
Whether the same is possible for the case of the FRW space-time is still an open question. 
It is definitely 
possible for the Milne Universe with the scale factor linearly growing with time.
This model was used many times to explain frame dependence of superluminarity (see, for example,
\citealt{zn1967, Page:1993mn}). In the recent
paper (\citealt{melia2012})
six different cosmologies allowing such a decomposition are described. 
In all these models a transformation to a stationary metric exists. 
As synchronous coordinates in a non-zero gravitational field 
are necessarily non-stationary, the non-stationarity of the frame
is the condition {\it sine qua non} for our description of the redshift. It is
not surprising, that the transformation to a stationary metric proposed by  \cite{melia2012} 
indeed allows us to reduce the cosmological 
redshift to other forms of redshift.
 
 However, these examples
to date do not even cover all FRW single-fluid scenarios, let alone more complicated
(though, more realistic!) models like the $\Lambda CDM$ model. So, we prefer to keep this
``tidal'' redshift
as a separate one, at least for reasonable practical purposes.

 We should also note that even in
the situation when such coordinate sets which allow decomposition of the
``tidal'' redshift into other
forms can be found for a particular FRW model, they include
absolutely different physical
values playing roles of ``distances'' and ``velocities''. As for these values in
the FRW frame,
despite the fact that they are formally not measurable, they still coincide (either directly or under a
simple re-definition) with observable entities: the proper distance at the
moment of emission coincides with the
angular distance, the rate of change of the angular distance is equal to the 
apparent velocity at the moment of emission
$v_{em}/(1+z)$, the proper distance now coincides with the proper motion
distance, etc. Ignoring peculiarities
of the tidal redshift in a comoving frame, and using a common intuition originating from more obvious
forms of redshift, we can come to ``evident'', but wrong conclusions.
Some examples will be described below.

\begin{figure}
\resizebox{0.9\hsize}{!}
{\includegraphics[]{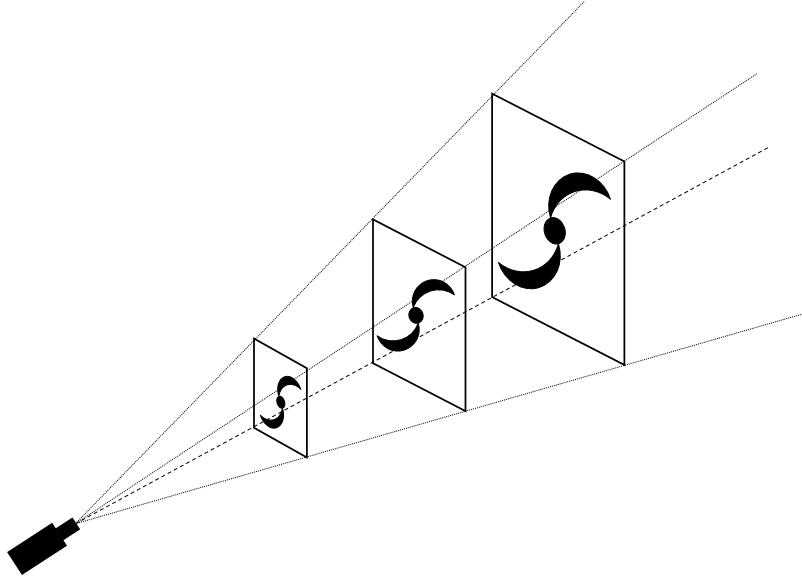}}
\caption{If in the flat universe we use a giant projector, then on all
parallel
screens images are non-distorted. Screens and the projector are moving
relative to each other due to cosmic expansion. However, the size of
an image on a given screen is determined at the moment of emission. If
we fill the universe with similar projectors transmitting the same
image, then on a given screen images of the same size can be produced
by faraway projectors and by near-by, similar to the situation with
angular sizes of galaxies at different distances in the sky.}
\label{homo}
\end{figure}

\subsection{Radar ranging in an expanding Universe}

According to the discussion above, 
a cosmological redshift is due to the non-stationarity of the metric used in
the derivation, but not due to a mysterious ``expansion of space''.
The ``expansion of space'' may be a good pedagogical concept, still, in general
it is not necessary to use it. Consider, for example, another effect which is
believed to support this idea. Namely, the difference in 
a radio signal travel time from an emitter to a reflector,
and back from the reflector to the emitter in an expanding Universe. The
return travel time is larger. 
Indeed, if we use notatation $t_{e}$, $t_r$, 
and $t_{obs}$ for the time of emission, reflection and observation of
a reflected signal, then using the equation for light propagation, 
we can see that the comoving distance
travelled by light is: 

\begin{equation}
\chi = \int_{t_e}^{t_r} \frac{dt}{a(t)} = \int_{t_r}^{t_{obs}} \frac{dt}{a(t)}.
\end{equation}

As the comoving coordinate of a reflector does not change, two integrals are equal. 
This means that for a monotonically
increasing $a(t)$ the return travel time $t_{obs}-t_r$ should be larger than the
forward travel time $t_r-t_e$.
This seems to support the expansion of space. 
Though,  \cite{chodor2007} argues that the moment of reflection
(in contrast to the moments of emission and observation of the reflected signal) is not directly
observable by the emitter (i.e. in the emitter's rest frame). 
So this moment is frame dependent. Moreover, it is possible to find a frame in which
this difference is completely explained by SR. 
This is correct, however, in the
present paper we want to stay in the FRW frame. Does this mean that we are
forced to accept the
expansion of space paradigm to reach an agreement with the calculated time intervals? Not at all!
Remembering that the velocity of light differs from $c$ for a distant observer in the
FRW frame,
we immediately see that on the way from the emitter to the reflector 
the velocity of the Hubble flow is
{\it added} to $c$, while on the way back it is {\it subtracted} from $c$, evidently resulting
in a larger time interval on the  way back. So, no additional concept (like
stretching space) is necessary to
explain the difference between these two time intervals.  

\subsection{Angular distance and its properties}

In this subsection we discuss why {\it angular} distances 
 in most of cosmological models starting from some $\chi$ 
decrease with increasing redshift. 

We remind the reader that the angular distance $D_{\theta}$ for an object with a
size $S$ which have
an angular diameter $\delta$ is by definition $D_{\theta}=S/ \delta$, so it is equal to the distance in Euclidean
geometry from which this object would have the same angular diameter. It is known that in
the FRW Universe 
$D_{\theta}=a_{em} \chi$.

It is important that a proper distance, $D=a\chi$, 
being by definition an entity which can not be measured
directly (such a measurement would require a chain of observers extending till the measured object,
organized in such a way that they make measurements in their vicinities at the same cosmic time $t$,
and after that all results should be summed, see \cite{weinberg1972}), if defined for the
moment of emission, has the same form ($D_{em}=a_{em} \chi$ if we put $\chi=0$ at the location of observer) 
as the angular distance.
The reason is simple: as radial rays are light geodesics, the expansion of the Universe
does not alter an angle between a pair of them. So, we see the object as it was at the time
of emission (of course, disregarding other properties like spectrum and luminosity which
are modified due to cosmic expansion).

We use the equation  for the light propagation in order to get distances and velocities
expressed through such a directly observable quantity as the redshift. Remembering that the redshift
$z$ obeys $(1+z)=a_{obs}/a_{em}$  we can rewrite the comoving coordinate of the
object seen now with the redshift $z$ as (details can be found in most of 
standard textbooks on cosmology):
 
\begin{equation}
\chi=\frac{c}{a(t_0)H_0} \int_0^z \frac{dz}{H(z)},
\end{equation}
where $t_0$ and $H_0$ are the present values of cosmic time and of the Hubble parameter.
For the FRW Universe filled with one type of
matter (fluid) with the equation of state $p=w\epsilon$ (where $\epsilon$
is density of the fluid), 
the time evolution
of the scale factor is $a \sim t^{2/(3+3w)}$, and correspondingly, $H(z)=H_0 (1+z)^{3(w+1)/2}$. This gives
the expressions:

\begin{equation}
\chi= \frac{c}{a(t_0)H_0}\frac{1}{1-\alpha}[(1+z)^{1-\alpha}-1],
\end{equation}
where we denote $\alpha=3(w+1)/2$.

Using this, we get the proper distance at $t_{em}$:

\begin{equation}
D=D_{\theta}=\frac{c}{H_0 (1-\alpha)(1+z)}[(1+z)^{1-\alpha}-1].
\end{equation}

We also can write down the expression for velocity: 

\begin{equation}
v=v_{em}=\frac{c}{1-\alpha} [1-(1+z)^{\alpha-1}].
\end{equation}

This is the well-known fact that $D_\theta(z)$ is not monotonic if $w>-1$, 
and has a maximum after which it
decreases, so that angular sizes increase with $z$. 
How can it happen? As radial rays are
light geodesics in the FRW coordinates any explanation based 
on the gravitational
focusing of light rays (see for example, \citealt{zn1967}) fails. 
Also, as our universe is flat, we cannot use
a popular explanation (see, for example, \citealt{mukhanov}) via light
propagation on a two-dimensional sphere, where due to curvature an observer
on a pole determines for objects behind the equator that $D_\theta$ is
decreasing while the physical (proper) distance is increasing. The failure of this
``explanation'' can be seen also from the fact that it should be applicable to de Sitter model
as well, however, $D_{\theta}(z)$ increases monothonically for $\alpha=0$.

The true reason of non-monotonic behavior of
$D_\theta$ is that two objects with the
same angular distances (which coincides with the proper distance at $t_{em}$) 
and different redshifts were indeed at the same distance 
from an observer when they emitted light visible now.
In a flat universe angles between light rays are invariant, i.e., a figure
formed by simultaneously emitted light pulses is transformed in a homothetic way during
light propagation in the expanding flat universe (see Fig.\ref{homo}). This
means that the angular distance is determined at the moment of emission, and
does not change by the cosmic expansion.
 
\begin{figure}
\resizebox{0.9\hsize}{!}
{\includegraphics[angle=90]{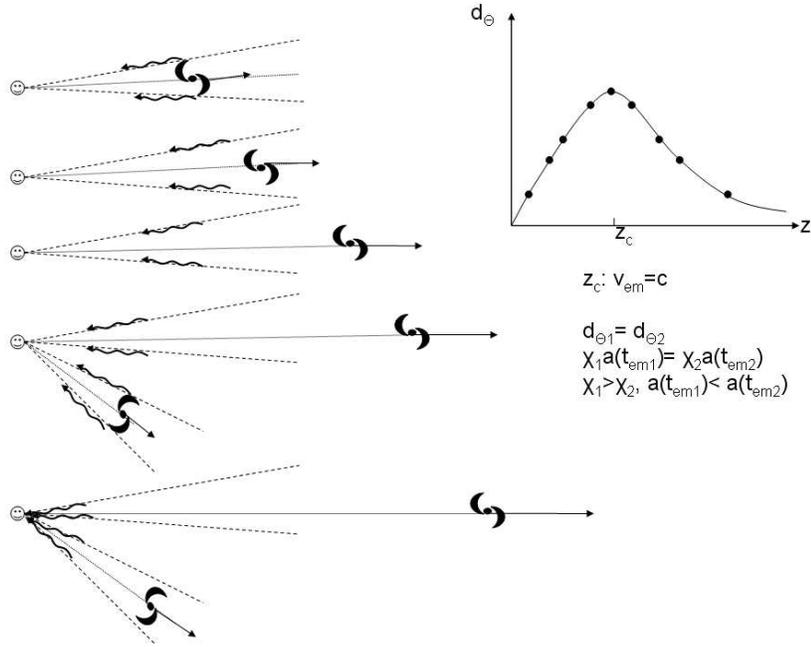}}
\caption{A cartoon illustrating the effect of decreasing of $D_{\theta}$ at
large redshifts, and the fact that $D_{em}$ is equal to the angular distance. 
Two galaxies have the same $D_{\theta}$ despite the fact that
they have different $z$. From Toporensky \& Popov (2014).} 
\label{fig_we}
\end{figure}

In Fig.\ref{fig_we} (reproduced from \citealt{tp2014}) we 
depicted schematically what happens. When the more redshifted object
was emitting, it receded
superluminally, so the light emitted towards us has been actually receding
from us. After some time when its $v$ diminishes due to decreasing $H$,
and became smaller than $c$, the light started to approach us.
At some moment the proper distance of the light again becomes equal to the
distance at which it had been emitted.

This illustration explains the pecularities of $D_{\theta}(z)$. 
It also tells us why the situation
in the de Sitter world (it corresponds to $w=-1$) is different. 
From eq. (6) we can see that $v$ is always
subluminal there, so this effect cannot appear. Also, it is evident  that $D_{\theta}(z)$  has its maximum
(when it exists) exactly when $v=c$ -- the fact that can be formally derived from eqs. (5-6).  

Finally, we would like to mention a curious fact that for 
the radiation-dominated Universe ($w=1/3$, and so $\alpha=2$) eq. (6) tells us
that the relation between recession velocities and cosmological redshifts has the
``naive'' form $v=cz$. We will discuss other peculiarities of the $w=1/3$
(radiation-dominated) Universe below, in Sec.4.

\section{Velocities}

This section is devoted to a discussion of several possible definitions of
the Hubble flow velocity. We also demonstrate how some of these velocities
evolve with cosmic time.

\subsection{Four types of velocities}

It is known that the Hubble flow can be characterized by several physical
parameters having the meaning of ``velocity'' of some kind.  
In the present section we compare four possible definitions.  

\begin{figure}
\resizebox{0.9\hsize}{!}
{\includegraphics{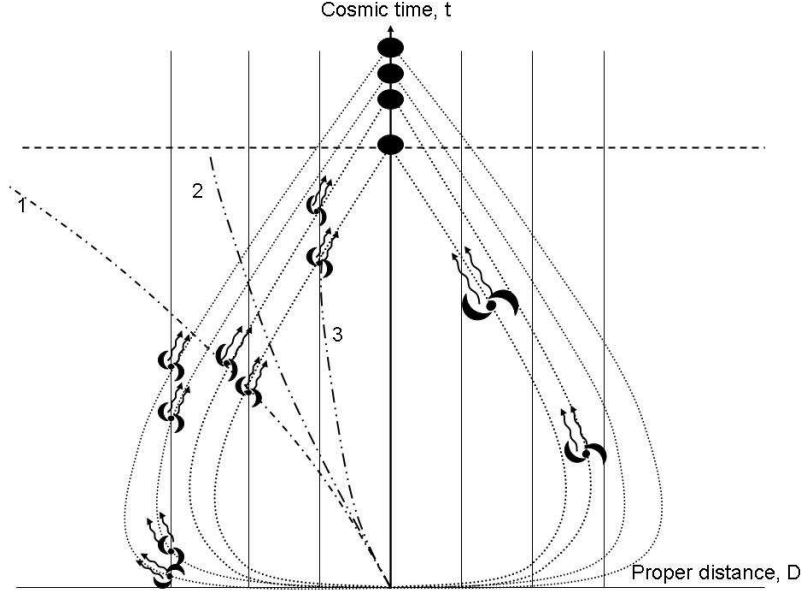}}
\caption{Dotted lines of the onion shape are light trajectories.
We observe objects on the light cone. Note that $D_{light}$ is
determined along the light cone. 
The plot illustrates observations  in
several consequent moments of time. Dashed-dotted curve ``1'' is the world line of a
galaxy in the Hubble flow. Vertical lines of constant proper distance
correspond to world lines of ``tethered'' galaxies. Dashed-dot-dotted lines ``2'' and ``3''
are world lines of galaxies with constant $D_{light}$.} 
\label{fig_sketch2014}
\end{figure}  

First of all, there are different
measures of ``distance'' used in different situations.  
In the preceding
sections we considered the proper distance at the time of emission --- $D_{em}$.
Similarly, we can consider the proper distance at the present time ---
$D_{now}$:  
\begin{equation}
D_{now}=a_{now} \chi = a_{em} (1+z) \chi = (1+z) D_{em}.
\end{equation}
This value is not directly measurable {\it per se}, though it is easy to
show that in the FRW Universe this value 
formally coincides with the proper motion distance and, thus,  in principle
can be measured.

Sometimes the third distance is used (especially in popular literature where it is
usually expressed in light years). It is the light travel distance ---
$D_{light}=ct$, where $t$ --- is the time during which the signal was
propagating.  It is determined along the light cone (see
Fig.\ref{fig_sketch2014}).
  
Obviously, in an expanding Universe, $D_{em}<D_{light}<D_{now}$. 
Taking time derivatives we obtain three
possible definitions of velocity with different properties.  
The velocity ``now'' is useful to form our mental image of the Universe
seen simultaneously as a whole (so-called ``God's view''): $v_{now}=\dot
D_{now}=\dot a_{now} \chi$.  
Oppositely, 
in the picture of the Universe seen by an observer, the velocity
at the time of emission $v_{em}=\dot D_{em}= \dot a_{em} \chi$ is more reliable. 
 
It is necessary to note,
however, that the apparent velocity measured by an observer 
differs from $v_{em}$ due to the difference in the march of time:
$v_{app}=v_{em}/(1+z)$. It is interesting that this velocity in the FRW Universe filled
with
matter with $w \le 1/3$ is always subluminal (see a recent duscussion and
references in \citealt{tp2014}). This, however, is a purely kinematical
effect and is not related to a rather special role of an ultra-relativistic matter
(having $w=1/3$) in modern physics. Indeed, the fact that this velocity is not restricted near
the Big Bang singularity for $w>1/3$ is a simple consequence of the fact that the value of $ \dot a/(1+z) \sim \dot a a$ for
the power law evolution $a\sim t^{1/\alpha}$ 
is not restricted near $t=0$ for $\alpha>2$. The law $a \sim t^{1/2}$  
is related to the equation of state $p=\epsilon/3$ only for the four-dimensional General Relativity (GR), and corresponds 
to other equations of state in modifications of GR, as well as in GR with
larger number of spatial dimensions. 

\begin{figure}
\resizebox{0.9\hsize}{!}
{\includegraphics{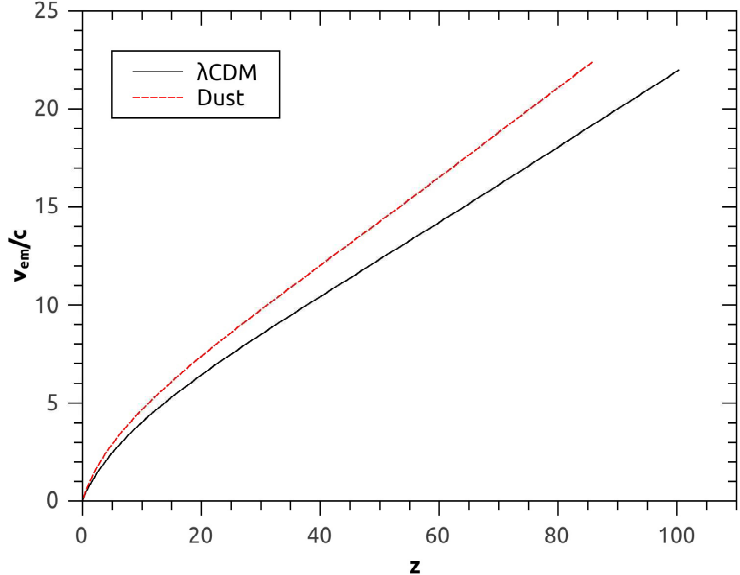}}
\caption{$v_{em}$ vs. $z$ in the dust and $\Lambda CDM$ models.} 
\label{fig_vemz}
\end{figure}  

Finally, the time derivative of the light travel distance 
gives us the velocity $v_l=dD_{light}/dt$. 
Simple geometric considerations show that the velocity $v_l$ is fully
determined by the redshift independently of the particular cosmological
model. Indeed, the difference in light travel
distance is obviously $dD_{light}=c (dt_{obs} - d t_{em})$. This equation,
after we substitute 
$1+z=dt_{obs}/d t_{em}$, immediately gives:
 
$$v_l=dD_{light}/dt_{obs}=c z/(1+z).$$  

In some papers and textbooks this value is referred to 
as an ``effective velocity of the Hubble flow''.  Indeed,
we can formally define velocity using a redshift by a non-relativistic formula
$v=cz$, and
apply the correction factor $(1+z)^{-1}$ to obtain
the apparent velocity measured by an observer. 
Then we derive the above mentioned equation.  
On the one hand, the definition of such an effective velocity uses formulae beyond their
range of application. On the other hand, the velocity $v_l$ is meaningful and
represents the time derivative of an appropriate distance.  
It is also evident that this velocity is always subluminal.  
So, why don't we consider
$v_l$ as a ``natural'' characteristic of the Hubble flow?  Formally this is possible,
however, there are significant drawbacks.  

\begin{figure}
\resizebox{0.9\hsize}{!}
{\includegraphics{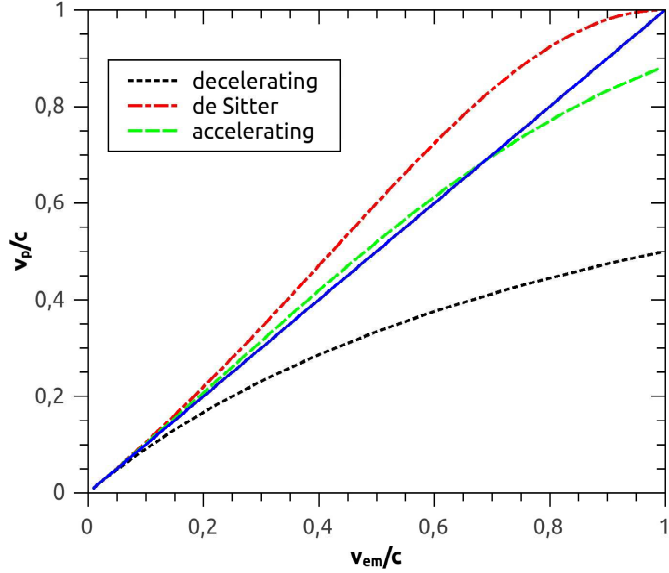}}
\caption{Relation between $v_p$ and $v_{em}$ for ``tethered'' galaxies in
different cosmologies. The straight solid line indicating $v_{em}=v_p$ is shown for convenience.} 
\label{fig_vz}
\end{figure}  

Remember, that the cosmological redshift is not determined at the moment of
emission, but is a cumulative effect, and, hence, the value of $v_l$ is determined by the
whole evolution history of the Universe since the time
of emission of the signal.  
To show how uncomfortable this property is with respect to our
intuitive notion of ``velocity'', let us consider a loitering
Universe (see \cite{sahni1992} and references therein).  
Such models received some attention and popularity some time ago, though
they require a rather exotic matter content.  Independently
of the exact reason to introduce loitering models, they are useful for our thought experiment.
Hence, consider a Universe with a scale factor which does not grow (at least
significantly) at some transitional epoch after which it continues
to grow rather rapidly (here we do not need to specify this rate in more 
detail).  What is the velocity of an object which emitted
light in the loitering epoch and is observed in the epoch of a rapid
expansion?  The velocity $v_{em}$ is evidently very small (since it
is proportional to $\dot a$ which is small during loitering).  The velocity
$v_l$ is, on the contrary, in general not small, because the loitering epoch
is characterized by objects with almost equal but not small, in general,
redshifts.  This discrepancy is even more pronounced if we allow the Universe
to contract a little during the loitering stage (though it is even
harder to achieve). In such situation $v_{em}$ is negative (since the scale
factor decreases at the moment of emission), though $v_l$ is positive (since this
contraction is compensated in the following epoch).

More formally, suppose we compare two different Universes so that each has in
its history two phases. The first phase in one Universe is  equivalent
to the first phase in another one. 
But the second phase of each Universe are drastically different.
An obvious example are dust and
$\Lambda CDM$ Universes, which share common early dust dominated stages, but
then they follow different evolutionary paths.
Suppose that two identical objects (one in each Universe, both have the
same age since the Big Bang) in the dust phases are observed by two
observers one of which lives in a $\Lambda$-dominated phase of one Universe,
and the second observer lives in the dust Universe.  Would they agree about the
recession velocity of the observed objects (we ignore practical
impossibility
for astronomers from different Universes to communicate and compare their results)?  It is
clear that they would determine the same $v_{em}$ which is
``recorded'' at the time of emission, and disagree about $v_l$ which
``encodes'' all the cosmic evolution from the time of emission to the time
of observation.

That is why we consider the velocity  $v_{em}$ (and its
observable ``twin'' $v_{app}$) as the most suitable parameter to
describe the Hubble flow at the moment of emission.  

There is, however, one important detail. If distant objects are formally marked with their
comoving coorinate $\chi$, the observers in two universes indeed
determine the same $v_{em}$ for sources emitting in the dust stage.  However,
real galaxies observed in a sky are not marked with $\chi$
(which should be calculated using a correct model of the Universe), but
instead marked with directly observable redshifts.  As redshifts 
depend on the whole evolution, two ``identical'' objects at the dust phase
observed much later would have different redshifts.  This means
that the values of $v_{em}$ as a function of $z$ are not the same in the common epoch of
these two universes  (see Fig.\ref{fig_vemz}).  The same is true if objects are
marked (by some clear evolution effects) with the time of emission: in a
given time of observation $t_{obs}$ the corresponding comoving coordinate
of an object emitting at $t_{em}$ depends on the expansion history.  So, only
$v_{em}$ as a function of $\chi$ has the same values
in the Universes with common evolution if observed later when the evolution
histories are different.  Nevertheless, as other functions
($v_{em}(z)$ or $v_{em}(t_{em})$) are obtained from $v_{em}(\chi)$ by an argument
re-definition, a receding object would always have a 
positive $v_{em}$, and vice versa. As we have seen above, this is not true
for $v_l$.

Finally, it is also worth noting that velocities defined directly through the time
derivative of scale factor obey the Hubble law since we have:
$$
v=\dot a \chi = \frac{\dot a} {a} a \chi= H D.
$$ 
Here we can take scale factors (and, correspondingly, the Hubble parameter)
either at the time of emission, or at the time of observation (i.e., ``now'').
On the other hand,  Hubble law is not valid  for the apparent velocity and $v_l$.

\subsection{``Tethered'' objects and the Hubble flow}

If we allow non-zero peculiar velocities for distant galaxies, the striking
difference between the sign of $z$ and the rate of proper distance 
changing can be found in the Universe without a non-trivial expansion
history.  Namely, as it has been pointed out by  \cite{davis2003} (see also \cite{clave2006}), 
it is possible to have blueshifted receding and redshifted approaching
objects.  The reason is that the resulting redshift of an object with
a peculiar velocity $v_p$ and recession velocity $v$ is given by
$(1+z)=(1+z_r)(1+z_p)$ where $z_r$ is redshift of the Hubble flow 
at the point where the observed galaxy is located (and is related to the
recession velocity at the emission by eq. (6)), and $z_p$ is the
Doppler shift due to the peculiar velocity which is calculated using the
standard relativistic formula

\begin{equation}
v_p=c\left(\frac{(1+z_p)^2-1}{(1+z_p)^2+1}\right).
\end{equation}

The difference in the presentations of eqs. (6) and (8) results, in particular, in a curious fact that
 so-called ``tethered'' galaxy with $v_p=-v$ 
(so that its proper distance does not change in time)
usually has a non-zero $z$.  It can be very easily understood in the case
of peculiar velocities close to the speed of light.  Since in most cosmological
models the redshift for a recession velocity equal to $c$ is finite, while
the blueshift for a peculiar velocity equal to $c$ obviously diverges, a
``tethered'' 
galaxy (with a constant proper distance) in the region of near-luminal recession
velocities will be blueshifted.  For a one-component perfect
fluid FRW model it is simple to show using eq. (6) that three different cases are
possible.  For the de Sitter Universe any ``tethered'' galaxy is redshifted
(remember that in contrast to other cases, in the de Sitter model $z \to
\infty$,
while $v_{em} \to c$).  For other accelerated Universes nearby
``tethered'' objects are redshifted, while distant are blueshifted.  Finally,
in decelerating models all ``tethered'' objects are blueshifted (see
Fig.\ref{fig_vz}).  

The same question can be considered for other definitions of velocity. What
happens
if we take the velocity $v_l$? The situation appears to be somewhat
different from 
the one described above.
 If we calculate the resulting redshift of an
object 
which has light travel time unchanged due to its non-zero peculiar velocity, 
we obtain $(1+z)=(1+z_{cosm})(1+z_{cl})(1+z_L)$. 
Here we denote by $z_{cosm}$ the cosmological redshift of an object in the
Hubble flow with zero $v_p$.
One can calculate that $v_p=cz/(1+z)$.
The first term in the r.h.s.
(the cosmological redshift) and the second term (the classical part of the
redshift
caused by peculiar velocity) represent time delay in the observer's frame,
while the third
term is caused by the Lorentzian time dilation in the emitter's frame. This
means that 
the first and the second terms for the object under consideration cancel out
(leaving
the observed light travel time unchanged), and the object is always
redshifted due
to the Lorentzian part of the peculiar velocity redshift.  

\section{Accelerations}

\subsection {Different acceleration measures}

In this section we discuss how an observer can see the Universe acceleration. As in the case of velocities,
there are several possible measures. In the present paper we consider only measures connected with time
derivatives of proper distance. Accelerations are denoted by the letter
$\mathfrak{A}$.

Similar to the case of velocities, in the ``God's view'' approach we can distinguish
between different measures --
an acceleration ``now'' and an acceleration at emission. Both are expressed by the same simple formula
$\mathfrak{A}_1=\ddot a \chi$ with the present-day scale factor for the former case
(acceleration ``now''), and the scale factor during emission for the latter. 

For the important case of a Universe filled by a single perfect fluid the acceleration
``now'' is equal to $cH_0[(1+z)^{1-\alpha}-1]$ while the acceleration at emission is $cH_0(1+z)^{\alpha}
[1-(1+z)^{\alpha-1}]$. Evidently, $\mathfrak{A}_1<0$ for $\alpha>1$ (decelerating Universe) and
$\mathfrak{A}_1>0$
for $\alpha<1$ (accelerating Universe). The acceleration at emission clearly diverges at a
horizon because $\ddot a$ diverges at the Big Bang.
An observer inside the Universe evidently has no access to these two types of acceleration.

What combination of cosmological variables is more reliable to describe what can
an observer see ``from inside''?
If we consider a rather clever observer who can calculate as well as to observe, we can imagine that such
an observer uses observational data in combination with an adequate model of expansion history of the Universe
in order to calculate  $v_{em}=\dot a(t_{em}) \chi$. 
Repeating this procedure after a short time interval 
the clever observer can get the difference in velocities at emission with
time, and calculate an acceleration
$\mathfrak{A}_2=dv_{em}/\Delta t_{obs}$, 
where $\Delta t_{obs}$ is the time interval measured by the observer. As usual, the ratio of time
intervals during emission and observation is equal to $1+z$, and we obtain 
$\mathfrak{A}_2=\ddot a (t_{em}) \chi /(1+z)$. For a one component Universe
$\mathfrak{A}_2=cH_0(1+z)^{\alpha-1}[1-(1+z)^{\alpha-1}]$. 
It is interesting that this acceleration tends to zero at the event horizon,
while diverges at the particle horizon.

It is possible also to consider $\mathfrak{A}_2$ 
from a different approach. Knowing the functional dependence of $v_{em}(z)$ and
time dependence of $z$ (\citealt{bq2007}): 
\begin{equation}
\dot z =H_0[1+z-H(z)/H_0]
\end{equation}
we easily get the observable rate of change of the velocity at emission:
\begin{equation}
 dv_{em}/dt=\frac{dv}{dz} \, \frac{dz}{dt} = \mathfrak{A}_2.
\end{equation}
 The second equality can be checked by straightforward calculation. 

However, for a ``not so clever observer'' who can only measure what is directly seen, 
$\mathfrak{A}_2$ is also unobservable.
A rate of the Universe expansion directly seen ``from inside'' 
is $v_{app}=\dot a(t_{em}) \chi /(1+z)$, and its time
derivative $\mathfrak{A}_3=dv_{app}/\Delta t_{obs}$ is not equal to
$\mathfrak{A}_2$. Moreover, the sign of $\mathfrak{A}_3$ can be different from 
the sign of $\mathfrak{A}_1$ and $\mathfrak{A}_2$ 
(obviously, as $\mathfrak{A}_1$ and $\mathfrak{A}_2$ differ by the always positive
multiplier
$1+z$, they have the same sign). For a one-component Universe with a perfect fluid
this can be easily seen
from eq. (6), which after dividing by 
$1+z$ gives us a functional dependence of $v_{app}(z)$. Again,  we can
write:
\begin{equation}
\frac{dv_{app}}{dt}=\frac{dv_{app}}{dz} \frac{dz}{dt}.
\end{equation}

As $\dot z>0$ for accelerating (in the usual sense) models, and remembering that $v_{app}$ grows for 
small $z$ and decreases starting from some $z_{cr}$, we get that
$\mathfrak{A}_3>0$ for $z<z_{cr}$ and $\mathfrak{A}_3<0$
for $z>z_{cr}$ in accelerating (from the ``God's view'') Universe. 
Similarly, we get that $\mathfrak{A}_3<0$ for $z<z_{cr}$,
and $\mathfrak{A}_3>0$ for $z>z_{cr}$ for deccelerating models with $\alpha<2$. 
As for $\alpha>2$ $v_{app}(z)$ is always increasing, $\mathfrak{A}_3<0$ for all redshifts in this case.

It can be easily seen also that $\mathfrak{A}_3$ always vanishes at the event horizon. 
As for its behavior at the particle
horizon, it depends on the equation of state. 
For $1<\alpha<3/2$ it tends to zero, for $\alpha>3/2$ it diverges (in the
boundary case of $\alpha=3/2$ it tends to the constant positive value equal to $c
H_0$ while $z$ tends to infinity).  
We see that $\mathfrak{A}_3$ has a more  complicated behavior in comparison with
$\mathfrak{A}_1$ and 
$\mathfrak{A}_2$.

Using the definition $1+z=a(t_{obs})/a(t_{em})$ we can write down a general expression
for $\mathfrak{A}_3$ through scale factors
(and their time derivatives) at the present time and at emission. Indeed, we can write
 $v_{app}=\chi a(t_{em})\dot a(t_{em})/a(t_{obs})$. Taking time derivatives and remembering that
$dt_{obs}=(1+z)dt_{em}$ we
obtain:
$$\mathfrak{A}_3=\frac{\chi}{a(t_{obs})^2}[\ddot a(t_{em}) a^2+\dot a(t_{em})^2 a - \dot
a(t_{em}) a(t_{em}) \dot
a(t_0)]$$.

Summarizing, we show that apart from the true (i.e. defined with respect to cosmic time intervals) acceleration 
``now'' (which is clearly belonging to the ``God's view'' picture), 
it is reasonable to introduce three different measures of
the cosmic acceleration -- the true rate of change of the recession velocity at emission
--- $\mathfrak{A}_1$, the apparent rate of
change of velocity at emission --- $\mathfrak{A}_2$, and the rate of change of the apparent velocity
-- $\mathfrak{A}_3$.

\subsection{Velocity evolution in realistic models}

 In this subsection we compare time evolution of velocities with time evolution of the redshift  
in models more complicated
then a one-fluid model.  We will see that an interesting
discrepancy between these two rates  appears in the $\Lambda CDM$ model, which is
currently believed to be the most appropriate to describe the
Universe we live in.  Let us recall what happens in a one-component model. 
The formula for time evolution of redshift (9) 
 immediately gives that redshifts become smaller in time in
decelerating Universes, and larger --- in accelerating Universes.
As for velocities, the situation is generally the same, and can be illustrated even
more easily. Since the comoving coordinate does not change in time,
time evolution of both $v_{em}=\dot a_{em} \chi$ and $v_{now}=\dot a_{now}
\chi$ is determined only by the second derivative
of the scale factor (at $t_{em}$ or ``now'').  The observed rate of change of
$v_{em}$ can be also found as:
$$\mathfrak{A}_2=\frac{dv_{em}}{dz} \frac{dz}{dt},$$
where the corresponding derivatives are calculated following eqs. (6) and (9).  Since
$dv_{em}/dz$ is always positive in a one-component Universe,
the sign of $\mathfrak{A}_2$ is determined by the sign of $\dot z$.

The situation changes for more general models including the important case of
the two-component $\Lambda CDM$.  It is evident that $v_{em}=
\dot a_{em} \chi$ decreases for objects which were emitting  when $\ddot
a_{em} <0$ (they have current redshifts larger than $z\sim 0.6$), and increases
for closer observed objects which were emitting when the Universe was already
accelerating.  On the other hand, using the exact formula for the
time evolution of the scale factor in the flat $\Lambda CDM$ Universe 
(see, for example, \citealt{ss2000}): 

\begin{equation}
a=(\mathrm{sinh}\,{(3/2)\sqrt{\Lambda/3}ct)}^{2/3},
\end{equation}
where $\Lambda=8\mathrm{\pi}G \rho_{vac}/c^2$ is the cosmological constant,
we can see that $\dot z > 0$ for $z$ less than approximately $2$. So that,
in our Universe (if the $\Lambda CDM$ model is a good 
approximation to reality) for $0.6 < z < 2$ redshifts increase while
corresponding velocities at emission decrease. 

Here we again come across 
the situation when the value of recession velocity (with its time
derivative) is determined at the time of emission, while the corresponding rate
of redshift changes depends on the whole expansion history till the moment of
observation (this manifests itself in an explicit dependence 
in the r.h.s. of eq.(9) on the {\it present} value of the Hubble parameter).

\begin{figure}
\resizebox{0.9\hsize}{!}
{\includegraphics{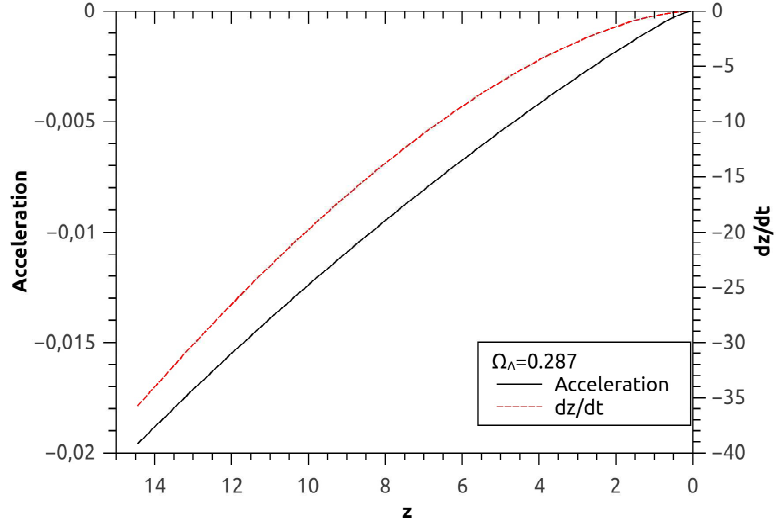}}
\caption{Evolution of apparent acceleration $\mathfrak{A}_2=(\dot v_{em}/c)/(1+z)$ 
(solid line) and $\dot z$ (dashed line). 
This plot corresponds to
an observer at the epoch when $\Omega_\Lambda=0.287$. We see that both
functions are always below zero.} 
\label{fig_omega1}
\end{figure}

\begin{figure}
\resizebox{0.9\hsize}{!}
{\includegraphics{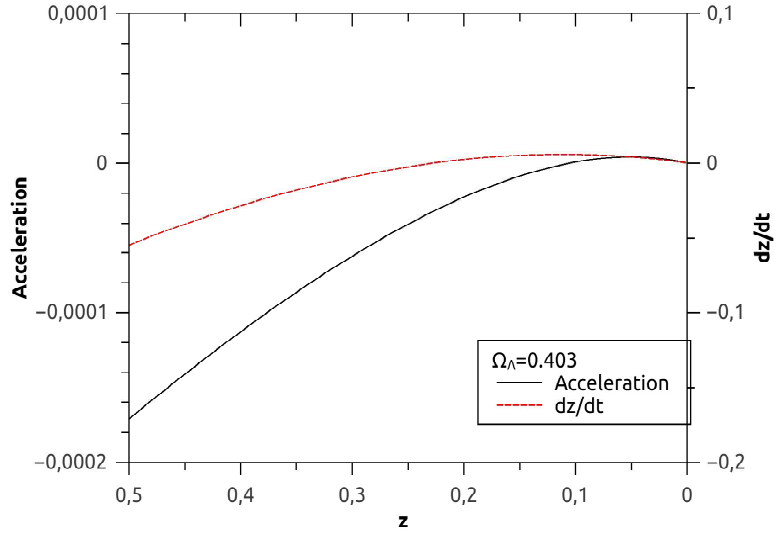}}
\caption{Evolution of apparent acceleration $\mathfrak{A}_2=(\dot v_{em}/c)/(1+z)$ (solid line) and $\dot z$ (dashed line). 
This plot corresponds to  
an observer at the epoch when $\Omega_\Lambda=0.403$.} 
\label{fig_omega2}
\end{figure}

\begin{figure}
\resizebox{0.9\hsize}{!}
{\includegraphics{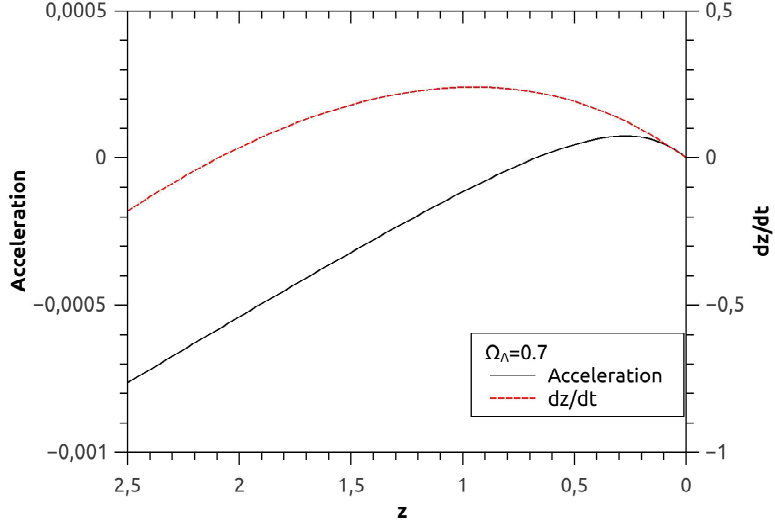}}
\caption{Evolution of apparent acceleration $\mathfrak{A}_2=(\dot v_{em}/c)/(1+z)$ (solid line) and $\dot z$ (dashed line). 
This plot corresponds to  
an observer at the epoch when $\Omega_\Lambda=0.7$. We see that at $z\sim 2$
$\dot z$ becomes positive.} 
\label{fig_omega3}
\end{figure}

\begin{figure}
\resizebox{0.9\hsize}{!}
{\includegraphics{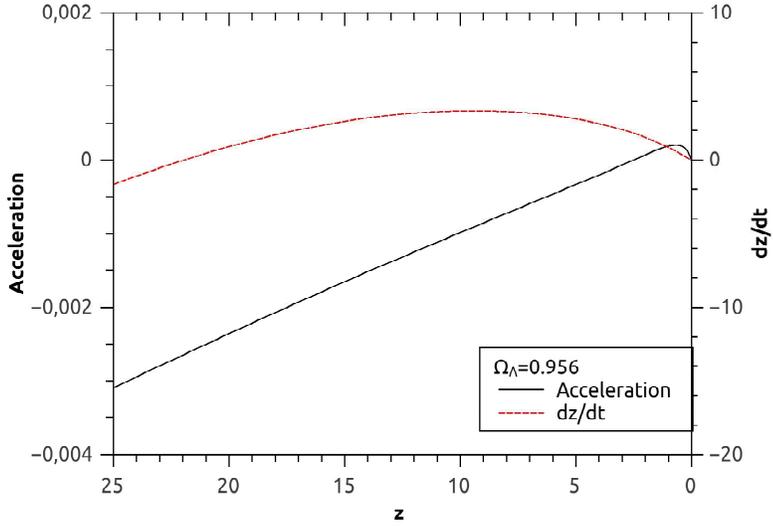}}
\caption{Evolution of apparent acceleration $(\dot v_{em}/c)/(1+z)$ (solid line) and $\dot z$ (dashed line). This plot corresponds to  
an observer at the epoch when $\Omega_\Lambda=0.956$.} 
\label{fig_omega4}
\end{figure}

\begin{figure}
\resizebox{0.9\hsize}{!}
{\includegraphics{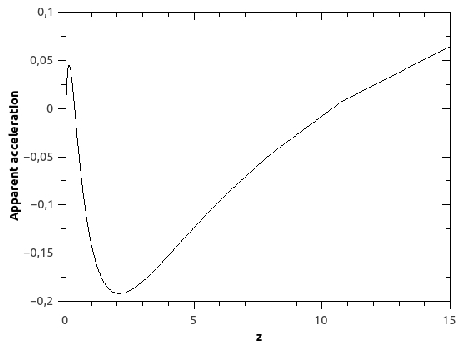}}
\caption{Apparent acceleration  $\mathfrak{A}_3(z)$ in the $\Lambda CDM$ model for the present day observer.} 
\label{fig_accel3}
\end{figure}

A question can arise: what is wrong with the equation (10) in the $\Lambda CDM$ case?  
Does this mean that $dv_{em}/dz$ changes its sign
for some range of $z$ (which would be rather unusual)?  No, it can be
checked that $dv_{em}/dz$ in the $\Lambda CDM$ model is always
positive.\footnote{In principle, it is possible to obtain $dv_{em}/dz<0$.
However, it requires exotic cosmological dynamics. For example, it happens
for the solution $H\sim t$ in $f(R)$-gravity for  $f(R)=R+R^2$.  }  
The true reason is that an analogue of formula (6) for a
multi-component models does not exist in the following sense: in eq.  (6)
the velocity is related to the corresponding value of the redshift
independently of the time of observation.  Obviously, we cannot expect
such property in a two-component model. In, say, $\Lambda CDM$ an observer
living in the dust-dominated epoch should use eq. (6) with $\alpha=3/2$, while
a ``later'' observer in the epoch of $\Lambda$-domination would use eq. (6) with $\alpha=0$.  Using
the exact formula (12)
it is possible to get a relation between $v$ and $z$ (at least
numerically). However, this result should include the time of observation
(or, equivalently, the parameter $\Omega_{\Lambda}$ which is {\it currently}
close to $0.7$ and is changing with time). 
For example, the parametrization $H(z)=H_0 (1+z)^{\alpha}$ now takes the
form: 
$$
H(z)=H_0(1+z)\left(1+\Omega_Mz+\Omega_{\Lambda}\left(\frac{1}{(1+z)^2}-1\right)\right)^{1/2}.
$$
 This means that eq. (10)
is not applicable in this case.  Some numerical results are shown in Figs.
\ref{fig_omega1}-\ref{fig_omega4}.
We can see that $v$ and $z$ both decrease for an observer
living at the epoch of deceleration.  After acceleration starts, nearby
objects have increasing $v$ and $z$. However, for a certain range
of intermediate redshifts we have $\dot v<0$ and $\dot z>0$.  This is again
because of the ``cumulative'' nature of redshifts and their time
derivatives along a light trajectory.
As for the rate of change of the apparent velocity, it has more complicated behvior with 
the redshift.  
It is clear that $\mathfrak{A}_3$ is positive for very small redshifts 
(at least in the present epoch when $\Lambda$ dominates) 
and tends to $cH_0$ for very large redshifts (shearing this asymptotic with
the pure dust model). The plot of $\mathfrak{A}_3(z)$ for the epoch when $\Omega_{\Lambda}=0.7$ 
is shown in Fig.\ref{fig_accel3}. It is visible in the figure that the
function changes sign twice, and it is negative for intermediate redshifts
(approximately from 1 to 10).

 For the realistic cosmological model $\dot z$ for different $z$ are
calculated in \cite{davis2004}. Observational prospects are discussed in
\cite{quer2012}. There is a hope that thanks to new large ground-based
telescopes and spectrographs it will be possible to measure $\dot z$ in the
next few decades. Variations of $v_{em}$ can be more elusive. 
At the moment the most precise method to determine angular distances is
related to maser measurements (see, for example, \citealt{kuo2013, hump2013} and references
therein). However, there is not much hope that changes in $D_\theta$ can be
detected in the near future.  

\section{Conclusions}

In this methodological note we presented a discussion of several
issues important in explanation of cosmological phenomenae.

In particular, we underline that in cosmology we mostly work in a
synchronous frame (typically, in the FRW frame), and properties of
such frames can be used to explain some non-trivial facts, such as
superluminal velocities in the Hubble flow, or Galilean summation of
velocities in cosmology. Interpretation of the cosmological redshift
also can be based on the properties of synchronous systems, and so,
formally, a very illustrative concept of ``stretching space'' is not
necessary to explain the origin of the redshift, if one intends to go
deeper into the nature of this phenomenon. Moreover,
a similar scheme (though more technically complex due to
spatial inhomogenity) can be used in other situations in GR 
where ``stretching of space'' cannot be considered as a useful concept (all we need
is a synchronous coordinate system). The cosmological redshift can be
reduced to gravitational effects in the FRW frame, however, not just to the well-known
gravitational time dilation.

In addition, we provided some illustrations related to different
velocities used in cosmology. 
We revisit the ``tethered galaxy'' problem and generalize results obtaned by
\cite{davis2003} for other definitions of recession velocities and other matter contents
(equation of state) of the Universe.

Finally, we discuss several possible measures of the cosmic acceleration.
Using them we compare time evolution of velocities and redshift in the 
$\Lambda CDM$ cosmology, and show why they evolve differently in this model.


We hope that these notes can be useful for better understanding of
some basic cosmological concepts by non-specialists. Especially, we
address this discussion to college lecturers who are teaching elements of
cosmology in their courses.

\begin{table}[t]
\caption{Basic parameters used in the paper}
\begin{tabular}{|l|p{3.1cm}|p{3.5cm}|p{4.5cm}|}    
\hline
 & & & \\
 $D_{now}$ & Proper distance now  &   $a(t_{obs})\chi$   &  $\frac{c}{H_0
(1-\alpha)}[(1+z)^{1-\alpha}-1]$ \\
 & & & \\
\hline
$D_{em}$  &  Proper distance at emission  &  $ a(t_{em})\chi$  & $\frac{c}{H_0 (1-\alpha)(1+z)}[(1+z)^{1-\alpha}-1]$ \\
 & & & \\
\hline
 $D_{light}$ &   Light travel distance & $c (t_{obs}-t_{em})$    &      $  \frac{c}{\alpha H_0}[1-(1+z)^{-\alpha}]$ \\
 & & & \\
\hline
 $v_{now}$ & Rate of change of the proper distance now & $\dot a(t_{obs}) \chi$ &
$\frac{c}{1-\alpha}[(1+z)^{1-\alpha}-1]$   \\
 & & & \\
\hline
 $v_{em}$ & Rate of change of the proper distance at emission  with respect to cosmic time &  $\dot
a(t_{em}) \chi$ &  $  \frac{c}{1+\alpha} [1-(1+z)^{\alpha-1}]$\\
 & & & \\
 \hline                                                                       
 $v_{app}$ & Apparent rate of change of the proper distance at emission  &  $\dot
 a(t_{em}) a(t_{em}) \chi / a(t_{obs})$ & $ \frac{c}{(1+\alpha)(1+z)}[1-(1+z)^{\alpha-1}]$\\
 & & & \\
\hline
 $v_{l}$ & Rate of change of $D_{light}$ &
$c\frac{a(t_{obs})-a(t_{em})}{a(t_{obs})}$ & $cz/(1+z)$  \\
 & & & \\
 \hline                         
$\mathfrak{A}_{now}$ & Acceleration now & $\ddot a(t_{obs}) \chi$ & $cH_0 [(1+z)^{1-\alpha}-1]$  \\
 & & & \\
\hline
$\mathfrak{A}_1$ & Rate of change of velocity at emission with respect to cosmic time & $\ddot
a(t_{em}) \chi$ &$ cH_0(1+z)^{\alpha}
[1-(1+z)^{\alpha-1}] $\\
 & & & \\
\hline
$\mathfrak{A}_2$ & Apparent rate of change of the velocity at emission & $\ddot
a(t_{em}) a(t_{em}) \chi /a(t_{obs})$ & $cH_0(1+z)^{\alpha-1}
[1-(1+z)^{\alpha-1}]$\\
 & & & \\
\hline
$\mathfrak{A}_3$ & Rate of change of the apparent velocity & 
$\frac{\chi}{a(t_{obs})^2}[\ddot a(t_{em}) a^2+\dot a(t_{em})^2 a - \dot
a(t_{em}) a(t_{em}) \dot a(t_{obs})]$ &   $\frac{cH_0}{1-\alpha}
[-(1+z)^{-1}-(\alpha-3)(1+z)^{\alpha-2}+(\alpha-2)(1+z)^{2\alpha-3}]  $   \\
 & & & \\
\hline
\end{tabular}
\end{table}

\vskip 1cm

\noindent
{\bf Acknowledgments}
We thank Sergey Pavluchenko and Alexander Petrov for comments on the manuscript.
S.P. in the Dynasty foundation fellow.


\bibliographystyle{apj}
\bibliography{bib_tp}

\end{document}